\documentclass[twocolumn,trackchanges]{aastex631}

\usepackage{xspace}

\shorttitle{The Aligned Orbit of WASP-193b}
\shortauthors{Yee et al.}

\graphicspath{{./}{figures/}}
\makeatletter
\def\input@path{{tables/}}
\makeatother

\defcitealias{Barkaoui2024}{B24}

\newcommand{\FitEcc}{\ensuremath{0.081^{+0.076}_{-0.055}}}
\newcommand{\FitMp}{\ensuremath{0.112^{+0.029}_{-0.034}}}
\newcommand{\FitRp}{\ensuremath{1.319^{+0.056}_{-0.048}}}
\newcommand{\FitRhop}{0.060 \pm 0.019}

\newcommand{\FitTeq}{\ensuremath{1250^{+22}_{-21}}}
\newcommand{\FitLambda}{\ensuremath{17^{+17}_{-16}}}
\newcommand{\FitMstar}{\ensuremath{1.018^{+0.064}_{-0.062}}}
\newcommand{\FitMstarB}{0.594 \pm 0.026}
\newcommand{\FitRstar}{\ensuremath{1.213^{+0.050}_{-0.042}}}

\newcommand{\FitAge}{\ensuremath{6.8^{+3.1}_{-2.4}}}

\newcommand{\Exofast}{\texttt{EXOFASTv2}\xspace}
\newcommand{\rmfit}{\texttt{rmfit}\space}

\newcommand{\bjdtdb}{\ensuremath{\mathrm{BJD}_\mathrm{TDB}}\xspace}
\newcommand{\Rstar}{\ensuremath{R_{\star}}\xspace} 
\newcommand{\Mstar}{\ensuremath{M_{\star}}\xspace}
\newcommand{\Rjup}{\ensuremath{R_\mathrm{J}}\xspace} 
\newcommand{\Mjup}{\ensuremath{M_\mathrm{J}}\xspace}
 
\newcommand{\Mearth}{\ensuremath{M_\oplus}\xspace}
\newcommand{\Rsun}{\ensuremath{R_\odot}\xspace} 
\newcommand{\Msun}{\ensuremath{M_\odot}\xspace}
\newcommand{\Rp}{\ensuremath{R_p}\xspace}
\newcommand{\Mp}{\ensuremath{M_p}\xspace}
\newcommand{\Teff}{\ensuremath{T_\mathrm{eff}}\xspace}
\newcommand{\Teq}{\ensuremath{T_\mathrm{eq}}\xspace}
\newcommand{\logg}{\ensuremath{\log{g}}\xspace}
\newcommand{\feh}{\ensuremath{\mathrm{[Fe/H]}}\xspace}
\newcommand{\vsini}{\ensuremath{v\sin{i}}\xspace}
\newcommand{\vmac}{\ensuremath{v_\mathrm{mac}}\xspace}

\newcommand{\secosw}{\ensuremath{\sqrt{e}\cos{\omega}}\xspace}
\newcommand{\sesinw}{\ensuremath{\sqrt{e}\sin{\omega}}\xspace}

\newcommand{\ms}{\ensuremath{\mathrm{m}\,\mathrm{s}^{-1}}\xspace}

\newcommand{\kms}{\ensuremath{\mathrm{km}\,\mathrm{s}^{-1}}\xspace}
\newcommand{\gcc}{\ensuremath{\,\mathrm{g}\,\mathrm{cm}^{-3}}\xspace}
\newcommand{\fluxunit}{\ensuremath{\mathrm{erg}\,\mathrm{s}^{-1}\,\mathrm{cm}^{-2}}\xspace}

\begin{document}

\title{The Super-Puff WASP-193\,b is On A Well-Aligned Orbit
\footnote{This paper includes data gathered with the 6.5 meter Magellan Telescopes located at Las Campanas Observatory, Chile}}

\author[0000-0001-7961-3907]{Samuel W.\ Yee}
\altaffiliation{51 Pegasi b Fellow}
\affiliation{Center for Astrophysics \textbar \ Harvard \& Smithsonian, 60 Garden St, Cambridge, MA 02138, USA}

\author[0000-0001-7409-5688]{Gudmundur Stef{\'a}nsson}
\affiliation{Anton Pannekoek Institute for Astronomy, University of Amsterdam, 904 Science Park, Amsterdam, 1098 XH}

\author[0000-0002-5113-8558]{Daniel Thorngren}
\affiliation{Department of Physics \& Astronomy, Johns Hopkins University, Baltimore, MD 21210, USA}

\author[0000-0002-0048-2586]{Andy Monson}
\affiliation{Steward Observatory, University of Arizona, 933 N. Cherry Ave., Tucson, AZ 85721}

\author[0000-0001-8732-6166]{Joel D.\ Hartman}
\affiliation{Department of Astrophysical Sciences, Princeton University, 4 Ivy Lane, Princeton, NJ 08544, USA}

\author[0000-0002-9003-484X]{David B. Charbonneau}
\affiliation{Center for Astrophysics \textbar \ Harvard \& Smithsonian, 60 Garden St, Cambridge, MA 02138, USA}

\author[0009-0008-2801-5040]{Johanna~K.~Teske}								% PFS Team
\affiliation{Earth and Planets Laboratory, Carnegie Institution for Science, 5241 Broad Branch Road, NW, Washington, DC 20015, USA}
\affiliation{The Observatories of the Carnegie Institution for Science, 813 Santa Barbara Street, Pasadena, CA 91101, USA}
\author[0000-0003-1305-3761]{R.~Paul~Butler}			% PFS Team
\affiliation{Earth and Planets Laboratory, Carnegie Institution for Science, 5241 Broad Branch Road, NW, Washington, DC 20015, USA}
\author[0000-0002-5226-787X]{Jeffrey~D.~Crane}			% PFS Team
\affiliation{The Observatories of the Carnegie Institution for Science, 813 Santa Barbara Street, Pasadena, CA 91101, USA}
\author{David~Osip}										% PFS Team
\affiliation{Las Campanas Observatory, Carnegie Institution for Science, Colina el Pino, Casilla 601, La Serena, Chile}
\author[0000-0002-8681-6136]{Stephen~A.~Shectman}		% PFS Team
\affiliation{The Observatories of the Carnegie Institution for Science, 813 Santa Barbara Street, Pasadena, CA 91101, USA}
% \author{Ian~Thompson}									% PFS Team
% \affiliation{The Observatories of the Carnegie Institution for Science, 813 Santa Barbara Street, Pasadena, CA 91101, USA}

%% PFS Observers -- awaiting reply
% \author{Sam Quinn}
% \affiliation{Center for Astrophysics \textbar \ Harvard \& Smithsonian, 60 Garden St, Cambridge, MA 02138, USA}
% \author{Yuri~Beletsky}
% \affiliation{The Observatories of the Carnegie Institution for Science, 813 Santa Barbara Street, Pasadena, CA 91101, USA}
% \author{Marcelo Said Tala Pinto}

%% Include?
% \author[0000-0002-4265-047X]{Joshua N.\ Winn}
% \affiliation{Department of Astrophysical Sciences, Princeton University, 4 Ivy Lane, Princeton, NJ 08544, USA}

\begin{abstract}

The ``super-puffs'' are a population of planets that have masses comparable to that of Neptune
but radii similar to Jupiter, leading to extremely low bulk densities
($\rho_p \lesssim 0.2\,\gcc$) that are not easily explained by standard core accretion models.
Interestingly, several of these super-puffs are found in orbits significantly
misaligned with their host stars' spin axes, indicating past dynamical excitation that may be
connected to their low densities.
Here, we present new Magellan/PFS RV measurements of WASP-193, a late F star hosting
one of the least dense transiting planets known to date
($\Mp = \FitMp\,\Mjup$, $\Rp = \FitRp\,\Rjup$, $\rho_p = \FitRhop\,\gcc$).
We refine the bulk properties of WASP-193\,b and use interior structure models to determine that the planet can be explained if it consists of roughly equal amounts of metals and H/He, with a metal fraction of $Z = 0.42$.
The planet is likely substantially re-inflated due to its host star's evolution, and expected to be actively undergoing mass loss.
We also measure the projected stellar obliquity using the Rossiter-McLaughlin effect, finding that WASP-193\,b is on an orbit well-aligned
with the stellar equator, with $\lambda = \FitLambda$ degrees.
WASP-193\,b is the first Jupiter-sized super-puff on a relatively well-aligned orbit, suggesting a diversity of formation pathways
for this population of planets.
\end{abstract}


\section{Introduction} \label{sec:intro}

Astronomical objects with sizes similar to that of Jupiter span several orders of magnitude in mass,
ranging from gas giant planets to small stars. The least massive of these are the ``super-puffs'',
extrasolar planets with masses between that of Neptune and Saturn
$20\,\Mearth \leq M_p \leq 100\,\Mearth$, corresponding to bulk densities of
$\rho_p \lesssim 0.2\,\gcc$.

These extremely low density planets are not easily explained by standard
core accretion models \citep{Pollack1996}, as their cores are thought to be susceptible to runaway gas
accretion, which would leave them with much larger gas envelopes and greater overall mass.
These planets may instead be the result of stalled or inefficient gas accretion
\citep{Lee2016,Lee2019} leading to masses much smaller than that of the typical gas giant.
Such objects may also be physically inflated due to tidal dissipation in the planet's interior
\citep{Millholland2020}, or their observed radii may be larger than expected due to dusty outflows
\citep{Wang2019a}.

% These puffy planets, owing to their large radii and atmospheric scale heights, are
% particularly amenable to detailed follow-up observations, which may help reveal their
% formation histories and internal structure.
Recent JWST transmission spectroscopy observations of WASP-107\,b indeed suggest a higher than expected internal heat flux \citep{Welbanks2024}, consistent with planet inflation due to heating by tidal dissipation.
Based on transit observations in the metastable He 10830 line, studies have also found that some of these Neptune-mass super-puffs are actively losing mass \citep[e.g.,][]{Spake2018,Vissapragada2024}.

Intriguingly, several of the lowest density Neptune-mass objects have been found to be on orbits severely misaligned with their host star's spin axes (Fig \ref{fig:density_obliquity}).
The three super-puff Neptunes with published obliquity measurements ($M_p < 100\,M_\oplus$, $\rho_p < 0.2\,\gcc$---WASP-107\,b \citep{Dai2017,Rubenzahl2021}, WASP-127\,b \citep{Allart2020}, and
KELT-11\,b \citep{Mounzer2022}--- are all on polar or retrograde orbits.
It has therefore been suggested that these planets are the outcomes of a formation pathway involving excitation onto a high-inclination, high-eccentricity orbit, followed by tidal circularization that may be
continuing to the present day \citep[e.g.,][]{Yu2024}.
Such a formation history would naturally explain their unusually high inclinations and inflated radii.

WASP-193\,b is a newly discovered transiting planet that is a member of the population of puffy
Neptunes. The planet was first detected by the WASP survey and described in
\citet[][hereafter B24]{Barkaoui2024}.
WASP-193\,b orbits a slightly evolved (\FitAge~Gyr old) solar mass star every 6.25 days, and has an inflated planetary radius of $R_p = 1.3\,R_J$.
\citetalias{Barkaoui2024} used radial velocity measurements from the HARPS and CORALIE spectrographs to measure the mass of WASP-193\,b and found that it has a bulk density of just $\rho_p = 0.06\,\gcc$,
making it one of the least dense planets discovered to date.

In this paper, we present new spectroscopic and photometric observations of WASP-193\,b,
including in-transit measurements taken with the goal of measuring the alignment between
the planet's orbit and stellar spin axis through the Rossiter-McLaughlin effect.
We describe the observations in Section \ref{sec:observations}, and our global re-characterization
of the system in Section \ref{sec:analysis}. We then discuss the implications of our measurements
for the internal structure and formation history of WASP-193\,b in Section \ref{sec:discussion},
and summarize in Section \ref{sec:conclusion}.

\section{Observations} \label{sec:observations}
\subsection{PFS Spectroscopy} \label{ssec:spectroscopy}

\begin{deluxetable}{cccc} \label{tab:rv_obs}
\tablecaption{PFS Radial Velocities for WASP-193}
\tablehead{
\colhead{Obs. Midpoint} & \colhead{Relative RV} & \colhead{$\sigma_\mathrm{RV}$} & \colhead{Exp. Time} \\
\colhead{\bjdtdb} & \colhead{[\ms]} & \colhead{[\ms]} & \colhead{[s]}
}
\startdata
% bjdtdb, rv, rverr, exptime
2460300.84220 & -5.39 & 6.99 & 600 \\
2460333.83057 & 7.53 & 5.20 & 600 \\
2460337.85463 & -18.99 & 5.62 & 600 \\
2460338.82387 & -0.29 & 5.78 & 600 \\
2460341.81969 & 2.59 & 5.22 & 600 \\
2460367.53478 & -14.13 & 6.87 & 900 \\
2460367.54523 & 6.74 & 5.38 & 900 \\
2460367.55428 & 0.00 & 6.91 & 600 \\
2460367.56132 & 16.18 & 6.23 & 600 \\
2460367.57589 & 44.94 & 5.60 & 600%
\enddata
\tablecomments{This table is published in its entirety in the machine-readable format.
A portion is shown here for guidance regarding its form and content.}
\end{deluxetable}

We observed the WASP-193 system with the Planet Finder Spectrograph (PFS;
\citealt{PFS_Crane2006,PFS_Crane2008,PFS_Crane2010}),
a high-resolution echelle spectrograph on the Magellan Clay telescope at
Las Campanas Observatory in Chile. All observations were made using the
0.3x2.5$^{\prime\prime}$ slit through an iodine cell, in 3x3 binning mode.
An additional high signal-to-noise template observation of WASP-193 was
taken without the iodine cell and used to derive precise radial-velocities (RVs).

We obtained a sequence of spectroscopic observations during a transit of
WASP-193\,b on the night of 2024 Feb 27. Observations began at 00 42 UT,
when the target was at airmass 1.9 and continued for 6.5 hours, covering
the entirety of the 4.25~hr transit event and an out-of-transit baseline comprising 0.75~hr pre-ingress and 1.5~hr post-egress. A total of
36 exposures were taken, with the first two having an exposure time of 900
seconds and the remainder being 600 second exposures. Conditions were clear
and stable throughout the sequence, with typical airmass-corrected seeing of
$0\farcs6$ or better. The spectroscopic data were reduced and precise RVs
extracted using a custom IDL pipeline described in \citet{PFS_Butler1996}.
The median instrumental RV precision achieved during the transit night was 5.5~m/s.

We also observed WASP-193 on 13 additional epochs between 2023 Dec 22 and
2024 Jun 02, with the goal of refining the spectroscopic orbit
of WASP-193\,b. These observations were taken in the same instrument
configuration and with 600s exposure times. The PFS RV data are presented
in Table \ref{tab:rv_obs}.

\subsection{Ground-Based Transit Photometry} \label{ssec:photometry}

In addition to the in-transit spectroscopic observations on 2024 Feb 27, we also obtained
simultaneous timeseries photometry. We observed with the Three-hundred MilliMeter Telescope (TMMT;
\citealt{TMMT_Monson2017}) and the 305 mm Las Campanas Remote Observatory
(LCRO) robotic telescope, both at Las Campanas Observatory in Chile, in
$I$ and $g^\prime$ bands respectively. We used AstroImageJ
\citep{AstroImageJ_Collins17} to perform aperture photometry and extract
flux time series for WASP-193. 
However, we noted that the TMMT data were affected by systematics on timescales similar to the transit ingress and egress duration, resulting in poor constraints on the transit timing.
As such, we only used the LCRO data in our final analysis.
Both datasets, together with variables used to detrend
the photometry during the transit fit, are available in machine-readable form
as Data behind the Figure for Figure \ref{fig:rm_plot}.

\subsection{Archival Data} \label{ssec:archival_data}

We made use of archival spectroscopic and photometric data in our analysis.
\citetalias{Barkaoui2024} published RV measurements from the HARPS and CORALIE
spectrographs. We included the HARPS RVs in our analysis but not the
CORALIE measurements, given their low precision (median instrumental uncertainty
$\sigma_\mathrm{RV,CORALIE} = 34\,\ms$) relative to the planet's RV
semi-amplitude ($K \approx 15\,\ms$).

\citetalias{Barkaoui2024} also published several follow-up transit light-curves obtained from the TRAPPIST-S \citep{TRAPPIST_Gillon2011} and SPECULOOS \citep{SPECULOOS_Jehin2018} facilities, dating back to 2015. We jointly fit these data in our analysis to extend the baseline of the transit data and precisely determine the transit ephemeris.

WASP-193 was observed by NASA's Transiting Exoplanet Survey Satellite (TESS;
\citealt{TESS_Ricker15}) in Sectors 9, 36, and 63, with the transits of
WASP-193\,b clearly detected in the TESS light curves. We obtained the 2-minute
cadence light curves reduced by the TESS Science Processing Operations Center
(SPOC; \citealt{TESS_SPOC_Jenkins2016}) for sectors 36 and 63 from MAST \citep{TESS_Sector_36,TESS_Sector_63}. These data were also previously analyzed by \citetalias{Barkaoui2024}.
% (\dataset[10.17909/x4v7-xd29]{http://dx.doi.org/10.17909/x4v7-xd29}, \dataset[10.17909/rzv6-r679]{http://dx.doi.org/10.17909/rzv6-r679})

In TESS Sector 9, WASP-193 was not one of the targets selected for 2-minute cadence
observations, and so was only observed in the 30-minute cadence full-frame
images (FFIs). While the TESS Quick-Look Pipeline (QLP; \citealt{TESS_QLP_Kunimoto2021})
produced a light curve for this object, we found that the planet's transit
depth in this sector is inconsistent with the depth measured in the
other TESS observations as well as with the ground-based transits presented by
\citetalias{Barkaoui2024}, potentially due to an incorrect dilution correction applied
to the QLP light curve.
We performed an independent extraction of a light curve using the procedures of the T16 project \citep{Hartman2025},
which performs image subtraction on the full frame images with the difference imaging pipeline originally developed for the Cluster Difference
Imaging Photometric Survey (CDIPS; \citealt{CDIPS_Bouma2019}).
As the transit depth measured from this light curve is consistent with the
other data, we used this reduction of the TESS Sector 9 data in our analysis.
This TESS light curve is provided in machine-readable form with Figure \ref{fig:tess_lc_plot}.
We detrended all TESS photometry by fitting the out-of-transit continuum to a basis spline, using the \texttt{Keplerspline} code
\citep{Keplerspline_Vanderburg2014,Keplerspline_Shallue2018}.
We then clipped the detrended light curves to use only data taken within
$\pm1.5$ transit durations of a transit event.

\subsection{Stellar Companion} \label{ssec:stellar_companion}

\begin{deluxetable}{lcc}
\tablecaption{Catalog Astrometry and Photometry for WASP-193 System \label{tab:comp_props}}
\tablehead{
    & \colhead{\textbf{WASP-193A}} & \colhead{\textbf{WASP-193B}}
}
\startdata
\textit{Identifiers} & & \\
Gaia DR3 ID & 5453063823882876032 & 5453063828179326976 \\
TIC ID 		& 49043968 & 49043967 \\
2MASS ID 	& J10572385-2959497 & J10572353-2959491 \\
WISE ID		& J105723.80-295949.7 & J105723.48-295951.1 \\
\hline
\textit{Astrometry} & & \\
Parallax (mas)\tablenotemark{a} & $2.648 \pm 0.015$ & $2.624 \pm 0.045$ \\
$\mu_{{\alpha}}$ (mas/yr)\tablenotemark{a} & $-49.055 \pm 0.012$ & $-49.494 \pm 0.038$ \\
$\mu_{{\delta}}$ (mas/yr)\tablenotemark{a} & $1.506 \pm 0.016$ & $2.259 \pm 0.049$ \\
RV (km/s)\tablenotemark{a} & $-2.78 \pm 0.82$ & -- \\
Ang. Sep. ($"$)\tablenotemark{b} & -- & 4.25 \\
Proj. Sep. (AU)\tablenotemark{b} & -- & 1600 \\
$\mathcal{{R}}_\mathrm{{chance}}$ \tablenotemark{b} & -- & $2.91 \times 10^{-7}$ \\
\hline
\textit{Photometry} & & \\
$G$ (mag)\tablenotemark{a} & $12.033 \pm 0.003$ & $15.936 \pm 0.003$ \\
$G_\mathrm{BP}$ (mag)\tablenotemark{a} & $12.335 \pm 0.003$ & $16.70 \pm 0.02$ \\
$G_\mathrm{RP}$ (mag)\tablenotemark{a} & $11.567 \pm 0.004$ & $14.926 \pm 0.006$ \\
$T$ (mag)\tablenotemark{c} & $11.634 \pm 0.008$ & $15.104 \pm 0.007$ \\
$J$ (mag)\tablenotemark{d} & 10.952 & 11.794 \\
$H$ (mag)\tablenotemark{d} & $10.810 \pm 0.034$ & $13.011 \pm 0.053$ \\
$K$ (mag)\tablenotemark{d} & $10.745 \pm 0.035$ & $12.804 \pm 0.048$ \\
W1 (mag)\tablenotemark{e}  & $10.640 \pm 0.026$ & $13.27 \pm 0.19$   \\
W2 (mag)\tablenotemark{e}  & $10.677 \pm 0.026$ & $13.18 \pm 0.19$   \\
W3 (mag)\tablenotemark{e}  & $10.491 \pm 0.093$ & $12.12$ 

\enddata
\tablerefs{\tablenotemark{a}Gaia DR3 \citep{GaiaEDR3_Brown2021,GaiaEDR3_Riello2021,GaiaEDR3_Lindegren2021,GaiaDR3_RVs_Katz2022};
\tablenotemark{b}Wide stellar binary catalog from \citet{GaiaEDR3_Binaries_El-Badry2021};
\tablenotemark{c}TESS Input Catalog \citep{TIC_Stassun2019};
\tablenotemark{d}2MASS \citep{TMASS_Cutri2003,TMASS_Skrutskie2006};
\tablenotemark{e}WISE \citep{WISE_Cutri2012}.
}
\end{deluxetable}

A search of archival survey data around WASP-193 revealed a faint, nearby star with
a sky separation of $4\farcs25$, also noted by \citetalias{Barkaoui2024}.
This star is identified in the \textit{Gaia}
\citep{GaiaDR3_Vallenari2022}, 2MASS \citep{TMASS_Cutri2003,TMASS_Skrutskie2006},
and WISE \citep{WISE_Cutri2012} surveys, and we list its photometric and astrometric
measurements in Table \ref{tab:comp_props}.\footnote{\textit{Gaia} parallaxes have been
corrected with the zero-point correction from \citet{GaiaEDR3_Lindegren2021}.}
Given this star has nearly identical parallax and
proper motion compared with WASP-193, the two components are almost certainly bound.
Indeed, the pair of stars appears in the \citet{GaiaEDR3_Binaries_El-Badry2021} catalog
of stellar binaries from Gaia EDR3, with a chance alignment statistic $\mathcal{R} = 2.9\times10^{-7}$,
indicating a high probability of being a bound pair with a sky-projected separation of
$\approx 1600$~AU. In the rest of this manuscript, we refer to this secondary stellar component
as WASP-193B.

\section{Analysis} \label{sec:analysis}

We performed a detailed analysis of the new and archival data to characterize the WASP-193 system.
We used the \texttt{SpecMatch-Synthetic} code to obtain stellar spectroscopic properties of the host star (\S\ref{ssec:spec_props}).
We fitted the broadband fluxes, transit photometry, and out-of-transit radial velocities with \texttt{EXOFASTv2} to retrieve the planetary bulk properties (\S\ref{ssec:global_analysis}).
Finally, we analyzed the in-transit RVs for the RM effect with the \texttt{rmfit} code (\S\ref{ssec:rm_analysis}).

\subsection{Spectroscopic Stellar Properties} \label{ssec:spec_props}

We derived spectroscopic stellar parameters for the planet host star WASP-193 by analyzing the high SNR, iodine-free template spectrum we obtained with PFS.
We used the \texttt{SpecMatch-Synthetic} code \citep{SpecMatchSynth_Petigura2015}, which
compares the observed spectrum with the \citet{Coelho2005} library of synthetic spectra, finding the best matches.
The code interpolates within the library stellar grid by taking linear combinations of the closest spectra to better match the target spectrum, and applies those coefficients to the grid properties to derive the final stellar parameters.
During this process, the library spectra are convolved with a broadening kernel to account for line broadening due to macroturbulence, assuming the $\Teff-\vmac$ relation from \citet{Valenti2005}, as well as stellar rotation, enabling a measurement of $\vsini$.

We determined that the planet host WASP-193 has $\Teff = 6000\pm60$~K,
$\logg = 4.02\pm0.08$~dex, $\feh = -0.12\pm0.06$~dex, and $\vsini=4.7\pm1.0$~\kms.
These results are all consistent within 1-$\sigma$ of the spectroscopic properties
derived by \citetalias{Barkaoui2024} from an analysis of HARPS and CORALIE spectra. 
We used the spectroscopic $\feh$ measurement as a prior for our global \Exofast
modelling (\S\ref{ssec:global_analysis}), and the \vsini measurement as a prior
for our analysis of the Rossiter-McLaughlin effect (\S\ref{ssec:rm_analysis}).

\subsection{Global System Characterization} \label{ssec:global_analysis}

\begin{figure*}
\includegraphics{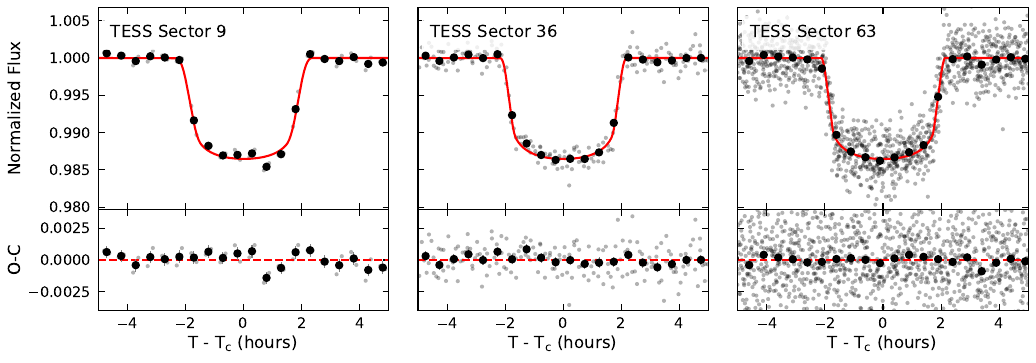}
\caption{TESS light curves of WASP-193 from Sectors 9, 36, and 63 (left to right),
phase-folded and clipped around the transit of WASP-193\,b. Solid black points
show measurements binned to 30 minute intervals. The red line shows the best-fitting
tranist model. The detrended TESS photometry used to create this figure are
available in machine-readable form.
\label{fig:tess_lc_plot}}
\end{figure*}

\begin{figure}
\includegraphics{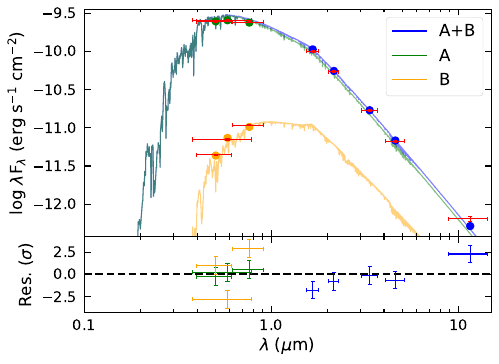}
\caption{Modelled two-component Spectral Energy Distribution (SED) for the WASP-193AB system, based on
the best-fitting stellar model from the \Exofast fit. The circles show the model
broad-band averages, while the red crosses denote observed fluxes from the Gaia, 2MASS,
and WISE catalogs. The horizontal error bars represent the filter bandwidths.
The lower panel shows the differences between the observed and modelled fluxes, colored
according to whether the observation represents the flux from WASP-193A (green),
WASP-193B (yellow), or a blend of both (blue).
\label{fig:sed_plot}}
\end{figure}

\begin{figure}
\includegraphics{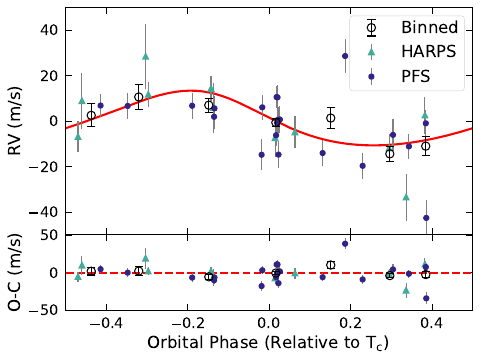}
\caption{Observed RVs of WASP-193 from HARPS and PFS, phase-folded to the orbital period
of WASP-193\,b, excluding those measurements taken during the planet's transit.
A relative RV offset between the two instruments has been removed.
The best-fitting RV model is plot as the red line, while open circles show the data
binned by orbital phase. \label{fig:rv_phased_plot}}
\end{figure}

% \begin{figure} \label{fig:ecc_hist}
% \includegraphics{ecc_hist.pdf}
% \caption{Posterior distribution for the eccentricity of WASP-193\,b, derived
% from our global modelling of transit photometry and radial velocities. The
% vertical dotted and dashed lines show the 68\% and 95\% upper limits on $e$.}
% \end{figure}

We used the \Exofast code \citep{ExoFASTv2_Eastman19} to self-consistently model the planet and stars in the WASP-193 system.
To constrain the planet properties, we fitted the out-of-transit RVs from PFS and HARPS, allowing for an independent RV offset for each instrument.
We parameterized the planet's orbital eccentricity in terms of $\secosw$ and $\sesinw$.
We also fitted the TESS and ground-based transit light curves.
% to perform a global characterization of the WASP-193 planetary system and
% derive uncertainties on the system parameters using a Differential Evolution-
% Monte Carlo Markov Chain (DE-MCMC) procedure.
While the TESS data are already corrected for dilution from nearby known stars within the same pixel, including from the secondary star WASP-193B, we allowed for an additional dilution term to account for potential uncertainties or errors in that correction.
We placed a Gaussian prior on this parameter with a width equal to 10\% of the existing correction.
For the ground-based photometry, we simultaneously detrended the light curves while fitting the transit model, using an additive detrending model.
The detrending vectors used were linear time, pixel position, and meridian flip for the LCRO photometry, while for the for the archival TRAPPIST-S and SPECULOOS data, we used the same variables as in \citetalias{Barkaoui2024}.

% We also modelled the LCRO ground-based photometry, fitting the
% transit model simultaneously with a linear detrending model against
% time, pixel position, and allowing for an offset due to a meridian flip.
% For the archival ground-based photometry from TRAPPIST-S and SPECULOOS, we detrended against the same variables as in \citetalias{Barkaoui2024}.

% In this analysis, we fitted the RV observations from PFS and HARPS, excluding those RV measurements
% taken during the transit of WASP-193\,b. The planet's eccentricity was allowed to vary,
% parameterized using \secosw and \sesinw.
% We fit for independent RV offsets between instruments, but did not include a
% long-term RV slope, as each instrument only observed WASP-193 for a limited
% baseline, and preliminary fits did not provide any evidence for such a trend.

In addition to modelling the planet, \Exofast simultaneously characterizes the stellar properties of the host star by
fitting the star's observed spectral energy distribution (SED) with the bolometric correction tables derived
from the MIST evolutionary models \citep{MIST0_Dotter2016,MISTI_Choi2016}.
This self-consistent approach incorporates the transit constraint on the stellar density and ensures the limb-darkening coefficients used in the transit model are consistent with the stellar properties, assuming a quadratic limb-darkening law and
coefficients tabulated by \citet{Claret2011,Claret2017}.
We accounted for both stars in the system using an updated version of \Exofast that allows fitting of a two-component SED.
As input data, we fitted the broadband catalog photometry from Gaia, 2MASS, and WISE (Table \ref{tab:comp_props}).
The $4\farcs3$ separation of the two stars is easily resolved by Gaia, but is close to the angular resolution limit of the 2MASS and WISE catalogs, so the fluxes reported for the secondary star in those catalogs may not be reliable.
We therefore fit the Gaia photometry for the two stars individually, while for the 2MASS and WISE photometry, we summed the measured fluxes and modelled the two stars as blended in those bands.
We excluded the 2MASS $J$-band flux in the
fit due to a flag indicating poor data quality.
% , while
% incorporating the stellar density constraint from duration of the planetary transit.
We placed Gaussian priors on the parallax of the star from Gaia DR3 \citep{GaiaEDR3_Lindegren2021}, \feh based on our spectroscopic analysis,
and an upper limit on the V-band extinction from the \citet{Schlafly2011} dust maps.
We required both stellar components to have the same distance, extinction, initial \feh and
age,\footnote{To account for potential systematic errors in the MIST stellar models,
we allowed stellar age of both stars to vary separately, but imposed a Gaussian prior on their age difference of width 0.1 Gyr.} since they are likely
a bound pair that formed together.
% The stellar model is also self-consistent with the limb-darkening parameters used
% to fit the time series photometry, .

% Given the presence of the close stellar companion WASP-193B, we used an updated version
% of \Exofast that is able to simultaneously model multiple stars.
% We required both components to have the same distance, extinction, initial \feh and
% age,\footnote{To account for potential systematic errors in the MIST stellar models,
% we allowed stellar age of both stars to vary, but penalized the model for age differences
% by an amount equivalent to a Gaussian prior of width 0.1 Gyr.} since they are likely
% a bound pair that formed together.
% We fit the Gaia photometry for the two stars individually, as Gaia is able
% to resolve binaries down to $\approx 1^{\prime\prime}$. However, the $4\farcs3$
% separation of the WASP-193AB pair is close to the spatial resolution of the
% 2MASS and WISE surveys. We therefore fit the 2MASS $HK$ and WISE W1, W2, and W3
% broadband photometry by summing the individually reported fluxes and modelling the
% photometry as blended in those bands. We excluded the 2MASS $J$-band flux in the
% fit due to a flag indicating poor data quality.

% \subsection{System Characterization Results}
\begin{deluxetable*}{lccc} \label{tab:exofast_results}
\tablecaption{Median values and 68\% Confidence Intervals for \Exofast fit of WASP-193}
\tablehead{\colhead{Parameter} & \colhead{Description} & \multicolumn{2}{c}{Value}}
\startdata
\\[-\normalbaselineskip]\multicolumn{2}{l}{Stellar Parameters:} & WASP-193A & WASP-193B \\
~~~~$M_\star$ ($M_\odot$) & Stellar mass & $1.018^{+0.064}_{-0.062}$ & $0.594 \pm 0.026$ \\
~~~~$R_\star$ ($R_\odot$) & Stellar radius & $1.213^{+0.050}_{-0.042}$ & $0.579^{+0.025}_{-0.026}$ \\
~~~~$\log{g_\star}$ (cgs) & Stellar surface gravity & $4.280^{+0.042}_{-0.050}$ & $4.686^{+0.030}_{-0.029}$ \\
~~~~$\rho_\star$ (g cm$^{-3}$) & Stellar density & $0.81 \pm 0.11$ & $4.31^{+0.51}_{-0.43}$ \\
~~~~$L_\star$ ($L_\odot$) & Stellar luminosity & $1.82^{+0.13}_{-0.12}$ & $0.0780 \pm 0.0046$ \\
~~~~$T_\mathrm{eff}$ (K) & Stellar effective temperature & $6080 \pm 130$ & $4007^{+70}_{-68}$ \\
~~~~$[\mathrm{Fe/H}]$ (dex) & Metallicity & $-0.108^{+0.055}_{-0.056}$ & $-0.027^{+0.066}_{-0.065}$ \\
~~~~$[\mathrm{Fe/H}]_0$ (dex)\tablenotemark{a} & Initial metallicity & $-0.035^{+0.051}_{-0.050}$ & $-0.035^{+0.051}_{-0.050}$ \\
~~~~Age (Gyr)\tablenotemark{a} & Stellar age & $6.8^{+3.1}_{-2.4}$ & $6.8^{+3.1}_{-2.4}$ \\
~~~~EEP & Equal evolutionary phase & $411^{+18}_{-28}$ & $315^{+11}_{-13}$ \\
~~~~$A_V$ (mag)\tablenotemark{a} & Visual extinction & $0.109^{+0.062}_{-0.070}$ & $0.109^{+0.062}_{-0.070}$ \\
~~~~d (pc)\tablenotemark{a} & Distance & $377.6^{+2.2}_{-2.1}$ & $377.6^{+2.2}_{-2.1}$ \\
\\[-\normalbaselineskip]\multicolumn{2}{l}{Planet Parameters:} & WASP-193A\,b \\
~~~~$P$ (days) & Period & $6.2463475 \pm 0.0000015$ \\
~~~~$T_c$ (BJD$_\mathrm{TDB}$) & Time of conjunction & $2459243.31080 \pm 0.00024$ \\
~~~~$R_P$ ($R_\mathrm{J}$) & Planet radius & $1.319^{+0.056}_{-0.048}$ \\
~~~~$M_P$ ($M_\mathrm{J}$) & Planet mass & $0.112^{+0.029}_{-0.034}$ \\
~~~~$\left(R_P / R_\star\right)^2$ & Planet-star area ratio & $0.01248^{+0.00028}_{-0.00027}$ \\
~~~~$K$ (m/s) & RV semi-amplitude & $12.3^{+3.1}_{-3.6}$ \\
~~~~$a$ (AU) & Semimajor axis & $0.0668 \pm 0.0014$ \\
~~~~$a/R_\star$ & Planet-star separation & $11.86^{+0.51}_{-0.58}$ \\
~~~~$i$ (deg) & Inclination & $89.25^{+0.51}_{-0.54}$ \\
~~~~$b \equiv a\cos{i}/R_\star$ & Transit impact parameter & $0.147^{+0.11}_{-0.099}$ \\
~~~~$e$ & Eccentricity & $0.081^{+0.076}_{-0.055}$ \\
~~~~$\omega$ (deg) & Argument of periastron & $42^{+84}_{-50}$ \\
~~~~$\sqrt{e}\cos{\omega}$ & Eccentricity vector & $0.12^{+0.22}_{-0.26}$ \\
~~~~$\sqrt{e}\sin{\omega}$ & Eccentricity vector & $0.12^{+0.13}_{-0.16}$ \\
~~~~$\rho_P$ (g cm$^{-3}$) & Planet density & $0.060 \pm 0.019$ \\
~~~~$\log{g_P}$ (cgs) & Planet surface gravity & $2.20^{+0.11}_{-0.16}$ \\
~~~~$T_\mathrm{eq}$ (K) & Planet equilibrium temperature & $1250^{+22}_{-21}$ \\
~~~~$\langle F \rangle$ (10$^9$ erg s$^{-1}$ cm$^{-2}$) & Incident flux & $0.547^{+0.039}_{-0.036}$ \\
~~~~$T_{14}$ (days) & Transit duration & $0.17733^{+0.00099}_{-0.00085}$ \\
~~~~$\tau$ (days) & Ingress/egress duration & $0.01821^{+0.00095}_{-0.00044}$ \\
\enddata

\tablenotetext{a}{These parameters are linked for both stellar components.}
\end{deluxetable*}

The best-fit stellar and planetary properties from our \Exofast analysis are presented
in Table \ref{tab:exofast_results}, while additional fit parameters, such as detrending
coefficients, are reported in the Appendix in Table \ref{tab:exofast_aux_results}.
Uncertainties in the parameters were determined by sampling the posterior distributions with a Differential Evolution-
Monte Carlo Markov Chain (DE-MCMC) procedure in the \Exofast code.
Figures \ref{fig:tess_lc_plot} -- \ref{fig:rv_phased_plot} show the best-fitting transit, stellar SED, and RV models.
We find that the primary stellar component, WASP-193A, is slightly evolved solar mass star ($\Mstar = \FitMstar\,\Msun$, $\Rstar = \FitRstar\,\Rsun$), with stellar
properties within 1-$\sigma$ of those reported by \citetalias{Barkaoui2024}. Meanwhile,
the secondary star WASP-193B is a late K star with a mass of $\Mstar = \FitMstarB\,\Msun$
(Figure \ref{fig:sed_plot}).

With the addition of new RV measurements from PFS, we obtain a smaller mass for the
transiting planet WASP-193\,b of $\Mp = \FitMp\,\Mjup$ (Figure \ref{fig:rv_phased_plot}),
approximately 1-$\sigma$ lower than the mass of $\Mp = 0.139 \pm 0.029\,\Mjup$ reported by
\citetalias{Barkaoui2024}. While these two results do not differ at a statistically
significant level, we note that measurements of the RV semi-amplitude $K$ tend
to be biased upward when the number of observations is small \citep{Shen2008,Piaulet2021},
so the downward revision in mass is not unexpected.

%% Include?
We also find a smaller radius for WASP-193\,b of $\Rp = \FitRp\,\Rjup$,
3-$\sigma$ lower than the $\Rp = 1.464\pm0.058\,\Rjup$ from \citetalias{Barkaoui2024}.
This discrepancy may arise from differences in the treatment of limb-darkening
between the two works, as the measurements of the transit depth from the TESS light
curves are similar (this work: $1.38\pm0.03\%$; \citetalias{Barkaoui2024}: $1.24\pm0.11\%$).
Still, because both the mass and radius of WASP-193\,b are smaller than previously thought,
we find a bulk density of the planet of $\rho_P = \FitRhop\,\gcc$, similar to the
\citetalias{Barkaoui2024} value, and WASP-193\,b remains an extremely low-density
super-puff planet.

One possible explanation for WASP-193\,b's inflated size is tidal heating from ongoing tidal circularization, as has been invoked to
explain the low densities of similar super-puffs like WASP-107\,b \citep{Piaulet2021}.
Dissipation arising from a relatively small but finite eccentricity of order $e \sim 0.05$, would provide sufficient internal heating to significantly inflate such planets, given their low masses and assuming Neptune-like tidal quality factors of $Q_P \sim 10^4$ \citep[e.g.,][]{Leconte2010,Welbanks2024}.
\citetalias{Barkaoui2024} reported that their data and analysis supported a
slightly eccentric orbit for WASP-193\,b, with $e = 0.056^{+0.068}_{-0.040}$.
With our additional PFS RV data, we measure $e = \FitEcc$, with a
68\% and 95\% upper limits of $e < 0.12$ and $e < 0.24$ respectively, consistent with this idea.
However, eccentricity measurements are subject to an upward bias when the true
eccentricity is low \citep{LucySweeney71}.
Using detailed simulations, \citet{Hara2019} estimated that the bias in eccentricity is $b_e \approx \sigma_\mathrm{RV} / K / \sqrt{\pi/(N_\mathrm{obs}-p)}$, where $p$ is the number of free parameters.
Based on the currently available data for WASP-193, the expected bias in the eccentricity could be as large as $b_e \sim 0.2$, indicating that nonzero eccentricity measurements up to 0.2 could be suspect.
Continued RV monitoring of the
system, or precise secondary eclipse timing measurements from facilities like JWST, would be needed to confirm any nonzero eccentricity.
If WASP-193\,b's orbit is indeed eccentric, an additional explanation must be sought to understand how it retained that eccentricity given its expected short tidal circularization timescale of $\tau_\mathrm{circ} \lesssim 0.2$~Gyr \citep{Adams2006}, many times shorter than the system age.
% Although WASP-193\,b has a relatively long orbital period, its low mass and
% large radius imply a short tidal circularization timescale of
% $\tau_\mathrm{circ} = \FitTcirc$~Gyr \citep{Adams2006}, assuming
% a planetary tidal quality factor of $Q_P = 10^6$.
% If the orbit of WASP-193\,b is really eccentric, heat from ongoing tidal
% interactions may help explain its inflated size, as has been invoked to
% explain the low densities of similar super-puffs like WASP-107\,b \citep{Piaulet2021}.
 % so continued, would be necessary to determine if this is the case.

% Such additional observations could also reveal the presence of an additional
% planet in the system, whose dynamical influence would be necessary to allow
% WASP-193\,b to retain a non-zero eccentricity over the $\sim 7$~Gyr age of the
% system.
% Although WASP-193\,b has a relatively long orbital period, its low mass and
% large radius imply a short tidal circularization timescale of
% $\tau_\mathrm{circ} = \FitTcirc$~Gyr, using Equation (3) of \citet{Adams2006}
% and a planetary tidal quality factor of $Q_P = 10^6$. While the latter
% quantity is poorly constrained, if WASP-193\,b is truly on an eccentric
% orbit, heat from ongoing tidal circularization may help explain its inflated
% size (\S\ref{ssec:tidal_inflation}).

\subsection{Rossiter-McLaughlin Analysis} \label{ssec:rm_analysis}

\begin{figure*}
\centering
\includegraphics[width=0.65\textwidth]{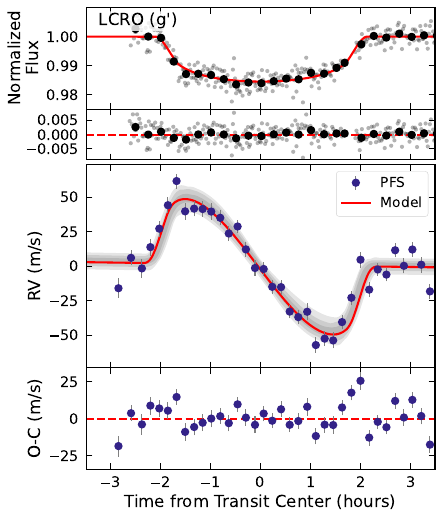}
\caption{In-transit photometric data from LCRO (top) and PFS RV measurements (bottom) for WASP-193,
taken on the night of 2024 Feb 27. The LCRO photometry was fit
as part of our global \Exofast analysis, and the transit constraints used as priors
for the \texttt{rmfit} analysis of the RM effect seen in the PFS RVs.
The red line shows the best-fitting model, and the gray shaded regions in the lower
panel show the 1, 2, and 3-$\sigma$ confidence intervals for the RM model.
The data behind this figure is available in machine-readable form.
\label{fig:rm_plot}}
\end{figure*}

\begin{deluxetable*}{lccc} \label{tab:rm_results}
\tablecaption{Priors and Fit Results for RM Analysis}
\tablehead{\colhead{Parameter} & \colhead{Description} & \colhead{Prior} & \colhead{Posterior}}
\startdata
$T_c$ (BJD$_\mathrm{TDB}$) & Transit midpoint & $\mathcal{N}(2460367.65298,0.00037)$ & $2460367.65277 \pm 0.00036$ \\
$P$ (days) & Orbital period & $\mathcal{N}(6.2463475,0.0000015)$ & $6.2463475 \pm 0.0000015$ \\
$\lambda$ (deg) & Sky-projected obliquity & $\mathcal{U}(-180.0,180.0)$ & $17^{+17}_{-16}$ \\
$v\sin{i_\star}$ (km/s) & Projected stellar rotation velocity & $\mathcal{N}(4.7,1.0)$ & $4.84^{+0.42}_{-0.35}$ \\
$i$ (deg) & Transit inclination & $\mathcal{N}(89.25,0.53)$ & $89.37^{+0.36}_{-0.45}$ \\
$R_p/R_\star$ & Planet-star radius ratio & $\mathcal{N}(0.1117,0.0012)$ & $0.1117 \pm 0.0012$ \\
$a/R_\star$ & Scaled semimajor axis & $\mathcal{N}(11.86,0.54)$ & $11.59^{+0.40}_{-0.42}$ \\
$u_1$ & Linear limb-darkening coefficient & $\mathcal{U}(0.0,1.0)$ & $0.38^{+0.29}_{-0.25}$ \\
$u_2$ & Quadratic limb-darkening coefficient & $\mathcal{U}(0.0,1.0)$ & $0.43^{+0.34}_{-0.30}$ \\
$\sqrt{e}\cos{\omega}$ & Eccentricity vector & $\mathcal{N}(0.12,0.24)$ & $0.07^{+0.16}_{-0.19}$ \\
$\sqrt{e}\sin{\omega}$ & Eccentricity vector & $\mathcal{N}(0.12,0.15)$ & $0.09^{+0.12}_{-0.13}$ \\
$\beta$ (km/s) & Intrinsic stellar line width & $\mathcal{N}(4.6,1.0)$ & $4.55^{+0.96}_{-0.94}$ \\
$K$ (m/s) & RV semi-amplitude & $\mathcal{N}(12.3,3.4)$ & $12.2 \pm 3.3$ \\
$\gamma$ (km/s) & PFS RV offset & $\mathcal{N}(-0.2,3.1)$ & $0.0 \pm 2.0$ \\
$\sigma_{\mathrm{RV}}$ (m/s) & PFS RV jitter & $\mathcal{U}(0.0,100.0)$ & $7.8^{+1.6}_{-1.5}$ \\
\enddata

\end{deluxetable*}

We measured the projected stellar obliquity, $\lambda$, by modelling the
 PFS RV timeseries from the night of 2024 Feb 27 with the \rmfit
code \citet{rmfit_Stefansson2022}\footnote{\url{https://github.com/gummiks/rmfit}}.
\rmfit models the Rossiter-McLaughlin (RM) anomaly
using the formulae of \citet{Hirano2010,Hirano2011},
using the standard parameterization
of the RM parameters in terms of $\lambda$ and $\vsini$.
We placed Gaussian priors on the planet's orbital parameters based on the output of our \Exofast analysis.
% and placed Gaussian priors on the transit and RV parameters
% derived from our \Exofast analysis. In particular, the transit midpoint at the
% epoch of observation is strongly constrained by our contemporaneous LCRO photometry along with the long baseline of TESS and ground-based data starting in 2015.
We also used a Gaussian prior on \vsini, from our \texttt{SpecMatch-Syn} analysis of the stellar spectrum (\S\ref{ssec:spec_props}).
We also repeated the RM analysis with an uninformative prior on \vsini and obtained consistent results, but elected to use the Gaussian prior, which helps eliminate a long tail toward high \vsini that is inconsistent with the spectroscopic line broadening measurement.
For the remaining parameters, we used uninformative priors,
and all priors are reported in Table \ref{tab:rm_results}.

\rmfit first finds the maximum likelihood model parameters using a differential evolution global optimization algorithm \citep{Storn1997} as implemented in \texttt{PyDE}.\footnote{\url{https://github.com/hpparvi/PyDE}}
We then sampled the posterior probability distributions through an MCMC analysis using the \texttt{emcee} code \citep{Emcee_Goodman10,Emcee_Foreman-Mackey13} to determine uncertainties on those parameters.
We confirmed the chains were well-mixed using the criteria for the Gelman-Rubin
statistic, $\mathrm{GR} < 1.01$ \citep{GelmanRubin}, and that the total chain length was $>$ 50 times the mean autocorrelation length.
The median and 67\% confidence intervals for each parameter are reported in Table \ref{tab:rm_results}.

The RV data, contemporaneous photometry and best-fitting model are plotted in Figure \ref{fig:rm_plot}.
We detect the RM anomaly at very high significance, with an RM semi-amplitude of
$\approx 50$~\ms, almost three times larger than the amplitude of the stellar reflex
motion from the orbiting planet.
We find that WASP-193\,b has a projected stellar obliquity of $\lambda = \FitLambda{^\circ}$,
consistent with a planetary orbit well-aligned with the stellar spin equator.

\section{Discussion} \label{sec:discussion}

\subsection{Interior Structure of WASP-193\,b} \label{ssec:context}

\begin{figure}
\includegraphics[width=\linewidth]{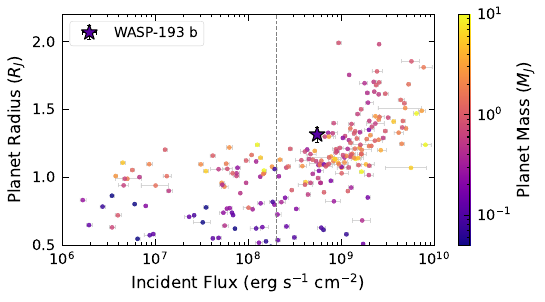}
\caption{Radii of known transiting exoplanets as a function of incident flux,
showing the onset of the hot Jupiter anomalous inflation at $\sim2\times10^8\,\fluxunit$
(vertical dashed line). Points are colored according to mass, and only planets with
masses measured to 33\% and radii to 20\% are shown here.
WASP-193\,b is marked with a star.
Data for literature planets are taken from the NASA Exoplanet Archive PSCompPars table
\citep{ExoplanetArchive_PSCompPars}.
\label{fig:radius_flux}
}
\end{figure}

\begin{figure}
\includegraphics[width=\linewidth]{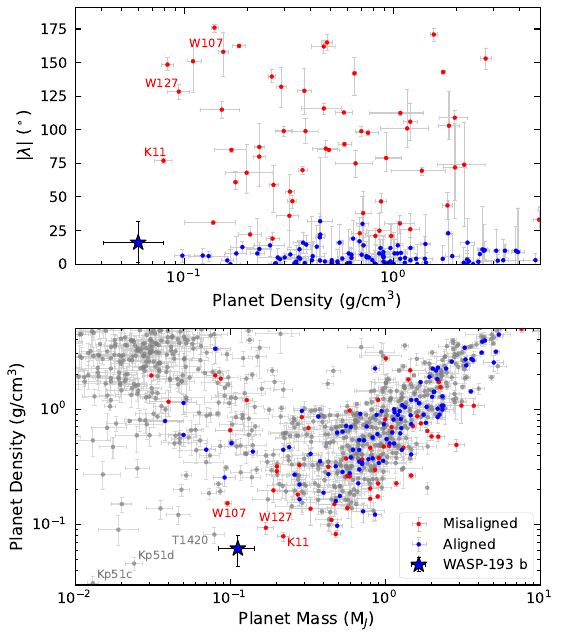}
\caption{
\textbf{Top:} Planets with measured projected stellar obliquity, as a function of bulk density. WASP-193\,b is the least dense planet with such a measurement, and is marked with a star. Planets with measured obliquities are shown in red if they are consistent with significant
misalignment with their host star's spin ($\lambda$ at least 1$\sigma$ greater than 10$^\circ$);
and in blue for well-aligned orbits (aligned to better than 10$^\circ$, roughly the scale at which our solar system is aligned).
\textbf{Bottom:} Masses and densities of known transiting exoplanets.
Objects with obliquity measurements are colored as in the top panel; those without such measurements are shown in gray.
% The planets most similar to WASP-193\,b 
The other known super-puffs are labelled
(WASP-107\,b: W107; WASP-127\,b: W127; KELT-11\,b: K11; TOI-1420\,b: T1420;
Kepler-51 c/d: Kp51c/d). Of these, only W107, W127, and K11 have obliquity measurements and all show significant misalignments.
In both panels, we use the same quality cuts
on mass and radius measurements as in Figure \ref{fig:radius_flux}.
Mass and density data from the NASA Exoplanet Archive, while obliquity measurements
are as compiled by \citet{Albrecht2022} and \citet{Knudstrup2024}.
\label{fig:density_obliquity}
}
\end{figure}

With a bulk density of $\rho = \FitRhop\,\gcc$, WASP-193\,b is one of the least
dense exoplanets known to date (Figure \ref{fig:density_obliquity}).
Only the Kepler-51 system \citep{Masuda2014} hosts planets with lower densities, although as a compact multi-planet system containing lower mass planets, Kepler-51 appears
qualitatively different from the systems containing isolated puffy Neptunes like WASP-107, TOI-1420, and WASP-193.

We applied interior structure models to WASP-193\,b to interpret its radius given its mass, equilibrium temperature, and age.  The planet has $\Teq = \FitTeq$~K, so we use the models of \citet{Thorngren2018} to account for the hot Jupiter inflation effect seen in objects with $\Teq \gtrsim 1000$~K (Figure \ref{fig:radius_flux}).
For the $\Teq$ of WASP-193\,b, the models call for approximately 1.1\% of the incident flux to be added as internal heating.  Accounting for the uncertainties in this value as well as the observed planet properties, we find that we can reproduce the mass and radius of the planet if it has a bulk metal content of $Z_p = 0.42\pm0.08$. 
Note that the quoted uncertainties do not include theoretical uncertainties on the modeling inputs, such as the H/He equation of state and the distribution of metals between the core and envelope. \citet{Thorngren2016} evaluated the potential effect of these uncertainties on the inferred properties and found that they may result in 10--20\% deviations, primarily arising from the choice of equation of state.

This inferred bulk metallicity is consistent with the predicted $Z_p = 0.42_{-0.19}^{+0.34}$ from the mass-metallicity relation of \citet{Thorngren2016}, although at these low masses, the distribution of predictions is broad.
Still, these models are able to explain the mass and radius of WASP-193\,b with a physically plausible interior structure and the ``standard'' hot Jupiter inflation, without needing to invoke an additional source of heat, for example due to tides.
The large size of WASP-193\,b is therefore consistent with the overall hot Jupiter population, once its low mass and high equilibrium temperature are accounted for.

The inferred metal content implies an upper limit on the planet's core mass $M_\mathrm{core} < M_p Z_p = 15.0^{+7.4}_{-6.5}\,\Mearth$, sufficiently large that standard core accretion theory predicts that it should have undergone runaway gas accretion.
Thus, WASP-193\,b appears to fall in the interesting and uncommon mass regime of giant planets that have avoided the fate of accreting a massive gas envelope, having roughly equal quantities of metals and H/He.

Of particular interest from an interior structure perspective is the fact that the host star appears to be near the end of its main sequence lifetime.
MIST stellar evolution tracks \citep{MISTI_Choi2016} indicate that WASP-193 has increased its total luminosity by a factor of $\sim2$ over the main sequence; as such, the planet spent most of its lifetime at an equilibrium temperature closer to 1000 K. 
As this is approximately the turn-on temperature for hot Jupiter inflation, the planet has had the opportunity to reinflate \citep{Lopez2016}, as we estimate the same planet would have been 20-30\% smaller at 1000 K, based on the minimum radius reached while evolving our planet interior structure models.  Further, the present-day radius of $\FitRp\,R_J$ is clearly inflated, so this planet is a valuable addition to the small list of planets showing direct signs of reinflation, along side such planets as  K2-132 b and K2-97 b \citep{Grunblatt2017}.

The low bulk density of WASP-193 b has an additional important implication; the planet is likely to be very vulnerable to runaway XUV-driven mass loss \citep[see e.g.][]{Thorngren2023}.
The XUV-luminosity tracks of \citet{Johnstone2021} predict that even at the star's advanced age, it is bombarding the planet with an XUV flux of $\sim3000$ erg s$^{-1}$ cm$^{-2}$.  Combined with the outflow models of \citet{Caldiroli2022}, we estimate a present-day mass loss rate $3.3 M_\oplus$ / Gyr.  The star will likely have had an XUV luminosity an order of magnitude higher in its first Gyr of life, so we expect that the planet has likely lost a substantial portion of its mass since formation, perhaps explaining its unusually high core-to-envelope ratio.

On the other hand, if the planet arrived more recently, such as through high-eccentricity migration (see Sec. \ref{ssec:formation_history}), the planet may not have lost as much mass as if it had always been on its present orbit.  The interaction between reinflation and XUV-driven mass loss will also be important in determining the history of this planet, but that would be a substantial project outside the scope of this article.  We suggest that future work on this planet begin with a measurement of the present-day stellar XUV luminosity as well as a search for signs of mass loss in the Helium triplet line during transit.

\subsection{Obliquity of WASP-193 b In Context} \label{ssec:formation_history}

Understanding how and when WASP-193\,b arrived at its present location would inform the interpretation of future mass loss measurements and whether the planet has lost a significant fraction of its gas mass.
Interestingly, the three other Neptune-mass super-puffs,
% of the other Neptune-mass giant planets with similar low densities and obliquity measurements,
namely WASP-107\,b \citep{Rubenzahl2021}, WASP-127\,b \citep{Allart2020}, and
KELT-11\,b \citep{Mounzer2022}, 
all have significantly misaligned orbits consistent with being polar or even retrograde to the stellar spin axis (Figure \ref{fig:density_obliquity}).
These observations are consistent with a dynamically excited migration history for those planets, perhaps involving 
% This evidence of past dynamical excitation is consistent with a relatively late migration of these planets to their current location,
eccentricity excitation via Kozai-Lidov oscillations followed by tidal circularization \citep[e.g.,][]{Yu2024}.

In contrast to other super-puffs, WASP-193\,b appears to be on a relatively well-aligned orbit.
We can confidently exclude the possibility that WASP-193 has a projected stellar obliquity as high as the $75-150^\circ$ measured for those objects.
While the true angle between the planet's orbital normal and the stellar spin axis, $\psi$, is not determined from the current data, a large true obliquity is \textit{a priori} unlikely \citep{Fabrycky2009}.
One question is whether this low obliquity is the result of tidal realignment between the stellar spin axis and the planetary orbit,
as has been proposed to operate for cool ($\Teff \lesssim 6200\,K$) planet-hosting stars with large convective envelopes \citep[e.g.,][]{Winn2010,Schlaufman2010}.
\citet{Attia2023} proposed a tidal efficiency factor
\begin{equation}
\tau \equiv \frac{M_\mathrm{conv}}{M_\star}\left(\frac{M_p}{M_\star}\right)^2\left(\frac{a}{R_\star}\right)^{-6};
\end{equation}
finding that for $\tau \lesssim 10^{-15}$, significant spin-orbit misalignments are observed for planets even around cool stars.
They therefore suggested $\tau \sim 10^{-15}$ as an approximate threshold for determining whether tidal realignment has operated.
In the case of WASP-193\,b, its low mass and wide orbit ($a/R_\star = 11.6$ given the current stellar radius) result in $\tau \approx 10^{-16}$, suggesting that its current low obliquity is unlikely to be due to post-formation realignment.

We found that the WASP-193 system does not have a large spin-orbit misalignment, unlike the other super-puffs.
This measurement excludes certain formation theories proposed to explain the preponderance of polar orbits particularly amongst Neptune-mass objects \citep{Albrecht2021,Knudstrup2024,Espinoza-Retamal2024}, such as disk-driven resonances that torque either the planet's orbit \citep{Petrovich2020} or the stellar spin axis \citep{Vick2023}, which drive the stellar obliquity to high values.
However, the precision of our measurement permits moderate misalignments of $\sim 30^\circ$, which would still be consistent with a high-eccentricity migration history driven by Kozai-Lidov oscillations.
Using \textit{Gaia} astrometry, we computed the angle between the sky-projected relative positions and velocities of the two stellar components \citep{Tokovinin1998,Tokovinin2016,Dupuy2022}, finding $\gamma = 43 \pm 3^\circ$. This indicates that the stellar binary orbit has a significant component in the sky-plane, in contrast to the orbit of the transiting planet, which is seen edge-on and would correspond to $\gamma = 0^\circ$.
Thus, the M dwarf stellar companion WASP-193B is a potential inclined perturber that could drive such dynamics.

Alternatively, WASP-193\,b may have formed without undergoing significant inclination excitation. 
The planet may have undergone early migration through a gas disk, or it may still have undergone
high-eccentricity migration but without inclination excitation, for example through secular interactions with a coplanar distant perturber \citep{Li2014,Petrovich2015}.
In the first case, the planet is unlikely to have any significant orbital eccentricity, and may even have close-by, non-transiting companions, as is more common for warm Jupiters \citep{Huang2016}.
We checked for Transit Timing Variations (TTVs) that may indicate the presence of nearby planets by fitting each observed transit independently, but found no evidence of any long-term variations above the typical $\sim$1--2 minute precision on the individual mid-transit times.
We do note that fitting the out-of-transit PFS RVs to a single Keplerian orbit required additional RV jitter larger than that expected for a star of this age \citep[e.g.,][]{Luhn2020a} as well as the excess noise measured over the hours-long timescale of the in-transit observations, so it is possible that the system does host additional short-period planets.
Given the limited time baseline of current data, it is not possible to determine how exactly the WASP-193 system may have formed, and why it has such a low stellar obliquity compared to the other super-puffs.
Continued RV monitoring or \textit{Gaia} epoch astrometry could reveal additional distant planetary companions, or remanant eccentricity from formation that has not yet been tidally damped.

% In the latter scenario where WASP-193\,b underwent coplanar high-eccentricity migration, an additional outer, coplanar planetary perturber must be present, as the known stellar companion's orbit is misaligned with the planet.
% These scenarios may be distinguished in the future by continued RV monitoring or through \textit{Gaia} epoch astrometry.

\section{Summary} \label{sec:conclusion}

We present new precise radial-velocity measurements of WASP-193 from Magellan/PFS, including
an in-transit time series clearly detecting the Rossiter-McLaughlin effect. We refine the radius
and mass measurement of the transiting planet WASP-193\,b, confirming its status as the lowest-density
Neptune-mass planets known to date.
Our interior structure models are able to reproduce the observed properties of WASP-193\,b, given its high equilibrium temperature and accounting for the hot Jupiter radius inflation effect. 
These models also indicate that the planet must have an unusually high bulk metal mass fraction of $\gtrsim 40\%$, and is likely experiencing significant mass loss.
Unlike other inflated Neptune-mass super-puffs,
the orbit of WASP-193\,b is comparatively well-aligned with its host star's spin axis, 
and this alignment is likely to be primordial.
This may suggest a diversity of formation pathways for these lowest density planets, or a common mechanism that produces both well-aligned and misaligned planets.
% disfavoring eccentricity excitation through the Kozai mechanism as the pathway by which this planet formed, although post-formation realignment cannot be definitively ruled out.

Continued RV monitoring of the system will improve constraints on WASP-193\,b's orbital
eccentricity, placing limits on the level of tidal heating, as well as potentially detect additional planets in the system that may shed light on the planet's formation. 
Given that the planet is likely losing mass, it is a prime target for detecting and measuring atmospheric outflows.
Its large atmospheric scale height also
means that WASP-193\,b is a particularly favorable target for characterization of its atmosphere.
Such observations would make for interesting comparisons with
other puffy but polar Neptunes to connect planetary structure, mass loss, and migration history.

\begin{acknowledgments}
This paper is based on observations obtained from the Las Campanas Remote Observatory that is a partnership between Carnegie Observatories, The Astro-Physics Corporation, Howard Hedlund, Michael Long, Dave Jurasevich, and SSC Observatories.

The authors thank the anonymous referee for helpful comments that improved the quality and clarity of this work.
S.W.Y. gratefully acknowledges support from the Heising-Simons Foundation.
S.W.Y. thanks Shreyas Vissapragada, Jason Eastman, Sarah Millholland, and W. Brennom for helpful conversations.
J.H. acknowledges support from NASA grant 80NSSC22K0409.

\end{acknowledgments}


\vspace{5mm}
\facilities{Magellan:Clay (PFS)}


\software{\texttt{astropy} \citep{Astropy13,Astropy18};
\texttt{EXOFASTv2} \citep{Exofast_Eastman13,ExoFASTv2_Eastman19};
\texttt{rmfit} \citep{rmfit_Stefansson2022}}



\appendix

Table \ref{tab:exofast_aux_results} contains additional fit parameters from our 
\Exofast analysis of the WASP-193 system.

\startlongtable
\begin{deluxetable*}{lcc} \label{tab:exofast_aux_results}
\tablecaption{Additional fit parameters from \Exofast analysis of WASP-193}
\tablehead{\colhead{Parameter} & \colhead{Description} & \colhead{Value}}
\startdata
\\[-\normalbaselineskip]\multicolumn{2}{l}{RV Parameters:} \\
$\gamma_\mathrm{HARPS}$ (m/s) & Relative RV offset & $-3175.4^{+3.0}_{-3.1}$ \\
$\sigma_\mathrm{jit,HARPS}$ (m/s) & RV jitter & $5.9 \pm 5.9$ \\
$\gamma_\mathrm{PFS}$ (m/s) & Relative RV offset & $-0.2 \pm 3.1$ \\
$\sigma_\mathrm{jit,PFS}$ (m/s) & RV jitter & $12.5^{+3.3}_{-2.6}$ \\
\\[-\normalbaselineskip]\multicolumn{2}{l}{Wavelength Parameters:} \\
$\delta_\mathrm{B}$ & Transit depth & $0.01708^{+0.00062}_{-0.00058}$ \\
$u_\mathrm{1,B}$ & Linear limb-darkening coeff & $0.554^{+0.044}_{-0.043}$ \\
$u_\mathrm{2,B}$ & Quadratic limb-darkening coeff & $0.222^{+0.040}_{-0.041}$ \\
$\delta_\mathrm{BB}$ & Transit depth & $0.01460^{+0.00046}_{-0.00044}$ \\
$u_\mathrm{1,BB}$ & Linear limb-darkening coeff & $0.300^{+0.043}_{-0.044}$ \\
$u_\mathrm{2,BB}$ & Quadratic limb-darkening coeff & $0.301 \pm 0.048$ \\
$\delta_\mathrm{g'}$ & Transit depth & $0.01674^{+0.00057}_{-0.00054}$ \\
$u_\mathrm{1,g'}$ & Linear limb-darkening coeff & $0.523 \pm 0.045$ \\
$u_\mathrm{2,g'}$ & Quadratic limb-darkening coeff & $0.271 \pm 0.050$ \\
$\delta_\mathrm{z'}$ & Transit depth & $0.01373^{+0.00036}_{-0.00035}$ \\
$u_\mathrm{1,z'}$ & Linear limb-darkening coeff & $0.187 \pm 0.033$ \\
$u_\mathrm{2,z'}$ & Quadratic limb-darkening coeff & $0.276^{+0.034}_{-0.033}$ \\
$\delta_\mathrm{TESS}$ & Transit depth & $0.01420^{+0.00035}_{-0.00033}$ \\
$u_\mathrm{1,TESS}$ & Linear limb-darkening coeff & $0.251^{+0.027}_{-0.026}$ \\
$u_\mathrm{2,TESS}$ & Quadratic limb-darkening coeff & $0.292 \pm 0.028$ \\
\\[-\normalbaselineskip]\multicolumn{2}{l}{Transit Parameters:} \\
$\sigma^2_\mathrm{TESS_{{9}}}$ & Added variance & $1.2^{+1.5}_{-1.1} \times 10^{-7}$ \\
$A_{D,\mathrm{TESS_{{9}}}}$ & Added dilution & $0.057^{+0.019}_{-0.020}$ \\
$F_\mathrm{0,TESS_{{9}}}$ & Baseline flux & $1.00016 \pm 0.00012$ \\
$\sigma^2_\mathrm{TESS_{{36}}}$ & Added variance & $-0.6^{+1.4}_{-1.3} \times 10^{-7}$ \\
$A_{D,\mathrm{TESS_{{36}}}}$ & Added dilution & $0.057^{+0.019}_{-0.020}$ \\
$F_\mathrm{0,TESS_{{36}}}$ & Baseline flux & $1.000509 \pm 0.000086$ \\
$\sigma^2_\mathrm{TESS_{{63}}}$ & Added variance & $-4.6^{+2.2}_{-2.1} \times 10^{-7}$ \\
$A_{D,\mathrm{TESS_{{63}}}}$ & Added dilution & $0.057^{+0.019}_{-0.020}$ \\
$F_\mathrm{0,TESS_{{63}}}$ & Baseline flux & $0.999877^{+0.000070}_{-0.000069}$ \\
$\sigma^2_\mathrm{LCRO}$ & Added variance & $2.95^{+0.60}_{-0.54} \times 10^{-6}$ \\
$F_\mathrm{0,LCRO}$ & Baseline flux & $0.99128 \pm 0.00021$ \\
$C_\mathrm{0,LCRO}$ & Detrending coefficient & $0.00866 \pm 0.00043$ \\
$C_\mathrm{1,LCRO}$ & Detrending coefficient & $-0.00249 \pm 0.00055$ \\
$\sigma^2_\mathrm{TRAPPIST_{{1}}}$ & Added variance & $1.007^{+0.086}_{-0.079} \times 10^{-5}$ \\
$F_\mathrm{0,TRAPPIST_{{1}}}$ & Baseline flux & $0.99673 \pm 0.00020$ \\
$C_\mathrm{0,TRAPPIST_{{1}}}$ & Detrending coefficient & $-0.00016 \pm 0.00035$ \\
$C_\mathrm{1,TRAPPIST_{{1}}}$ & Detrending coefficient & $0.00229 \pm 0.00059$ \\
$\sigma^2_\mathrm{TRAPPIST_{{2}}}$ & Added variance & $1.091^{+0.071}_{-0.067} \times 10^{-5}$ \\
$F_\mathrm{0,TRAPPIST_{{2}}}$ & Baseline flux & $1.00775 \pm 0.00031$ \\
$C_\mathrm{0,TRAPPIST_{{2}}}$ & Detrending coefficient & $-0.00008 \pm 0.00041$ \\
$C_\mathrm{1,TRAPPIST_{{2}}}$ & Detrending coefficient & $-0.00669 \pm 0.00057$ \\
$\sigma^2_\mathrm{TRAPPIST_{{3}}}$ & Added variance & $6.02^{+0.89}_{-0.83} \times 10^{-6}$ \\
$F_\mathrm{0,TRAPPIST_{{3}}}$ & Baseline flux & $1.00066 \pm 0.00028$ \\
$C_\mathrm{0,TRAPPIST_{{3}}}$ & Detrending coefficient & $0.0032 \pm 0.0017$ \\
$C_\mathrm{1,TRAPPIST_{{3}}}$ & Detrending coefficient & $0.0054 \pm 0.0018$ \\
$\sigma^2_\mathrm{TRAPPIST_{{4}}}$ & Added variance & $-2.75^{+0.13}_{-0.11} \times 10^{-5}$ \\
$F_\mathrm{0,TRAPPIST_{{4}}}$ & Baseline flux & $0.99878^{+0.00034}_{-0.00033}$ \\
$C_\mathrm{0,TRAPPIST_{{4}}}$ & Detrending coefficient & $0.00105^{+0.00082}_{-0.00083}$ \\
$C_\mathrm{1,TRAPPIST_{{4}}}$ & Detrending coefficient & $0.00047^{+0.00098}_{-0.00100}$ \\
$\sigma^2_\mathrm{SPECULOOS}$ & Added variance & $5.25^{+0.66}_{-0.61} \times 10^{-6}$ \\
$F_\mathrm{0,SPECULOOS}$ & Baseline flux & $0.99558 \pm 0.00018$ \\
$C_\mathrm{0,SPECULOOS}$ & Detrending coefficient & $0.01094 \pm 0.00069$ \\
$C_\mathrm{1,SPECULOOS}$ & Detrending coefficient & $-0.0219 \pm 0.0013$ \\
\enddata

\tablecomments{The transit parameters for TRAPPIST and SPECULOOS refer to the follow-up photometry obtained by \citetalias{Barkaoui2024} and re-fit in our analysis, using the same detrending parameters. The four TRAPPIST light curves are indexed in chronological order.}
\end{deluxetable*}

\bibliography{software,instruments,catalogs,bibliography}{}
\bibliographystyle{aasjournal}



\end{document}